\documentclass[onecolumn,aps, preprint ,showpacs]{revtex4}
\usepackage{graphicx}
\usepackage{epsfig}
 \begin{document}
\setlength{\unitlength}{1mm}
\bibliographystyle{unsrt} 
\title{ Spatial Constraint   Corrections to the  Elasticity of dsDNA   Measured  with  Magnetic Tweezers.
}
 \author{C. Bouchiat}
\affiliation{Laboratoire de Physique Th\'eorique de l'Ecole Normale Sup\'erieure \\ 
24, rue Lhomond, F-75231 Paris Cedex 05, France.}
\date {\today}
\begin{abstract}
In the present  paper, we have studied a discrete  version of the WLC model, which incorporates the spatial constraints 
imposed by the magnetic  tweezer,  used in current micro-manipulation experiments.
 These obstruction effects are  relevant for ``short" molecules,
involving  about two thousand base pairs or less. Two elements 
 of the device have to be considered: first, the fixed  plastic slab on which is stuck  one molecule end, second, a   magnetic 
bead which is used to pull (or twist) the attached  molecule free end. 
 We have developed quantitative  arguments  showing that the bead surface can be replaced 
by its tangent plane at the anchoring point, when it is close 
 to the bead  south pole relative  to the pulling  direction.
We are, then, led to a  confinement model  involving  two repulsive plates: first,  the fixed anchoring plate, second, a  fluctuating 
plate,  simulating the bead,  in thermal equilibrium with the attached molecule and the ambient fluid. 
The  bead obstruction effect reduces to  a slight upper shift of the elongation, about four times smaller and with the same sign as
 the effect induced by the anchoring plate. This result, which may contradict  naive expectations, has been 
qualitatively confirmed within  the soluble  ``Gaussian" model for flexible polymers. A study of the  molecule 
elongation versus the contour length $L $ exhibits a significant non-extensive  behavior.
Although the  curve  for ``short" molecules is well fitted by a straight line,  with its slope very close to
the  prediction of the standard WLC
model, it does not pass  through the origin, due  the presence of an offset term independent of $L$. 
This  leads to  a $ 15 \% $ upward shift 
of the elongation for a  2 kbp  molecule. Finally, the need  for  thorough  analysis of the  spatial 
constraints  in   super-coiled dsDNA  elasticity measurements   is   illustrated by  ``hat" curves, 
 giving  the elongation versus the torque.
 \end{abstract}
\maketitle

%%%%%%%%%%%%%%%%%%%
  \newcommand \be {\begin{equation}}
\newcommand \ee {\end{equation}}
 \newcommand \bea {\begin{eqnarray}}
\newcommand \eea {\end{eqnarray}}
\newcommand \nn \nonumber
\def \(({\left(}
\def \)){\right)}
 \def \vr{{\mathbf{r}}}
\def \vv{{\mathbf{v}}}
 \def \vk{{\mathbf{k}}}
\def \vq{{\mathbf{q}}}
\def \vf{{\mathbf{f}}}
\def \vt{{\mathbf{t}}}
\def \vu{{\mathbf{u}}}
\def \vp{{\mathbf{p}}}
\def \vR{{\mathbf{R}}}  
 \def \va{{\mathbf{a}}}
 \def \vb{{\mathbf{b}}}
\def \vF{{\mathbf{F}}}
\def\vr{{\mathbf{r}}}
\def\vt{{\mathbf{t}}}
\def\vz{{\mathbf{z}}}
\def\vrp{{\mathbf{r_{\perp}}}}
\def\vtp{{\mathbf{t_{\perp}}}}

\def\bra{\langle}
\def\ket{\rangle}
%%%%%%%%%%%%%%%%%% 
\section{Introduction}
    
In the last decade, single particle biophysics has developed 
into a  very active field of research \cite{busta,allem,smith,perk,strick}. 
In particular, micro-manipulation experiments are now recognized as  valuable  tools
to observe, in  real time, a single  double-strand DNA (dsDNA) molecule  interacting   with  the 
proteins involved in the cell duplication processes. The basic principle  is to look for 
   sudden  variations of the stretched  dsDNA molecule elongation, which occur  when
 the biochemical reaction is taking   place.
(For two recent reviews see  the references \cite{busta,allem}.)
 
 In recent  experiments  \cite{Terence,Ebri},  there is a tendency  to
use relatively  short   segments with 2000 base pairs (2 kbp), corresponding to  $ L \simeq 680 $~nm; this number  is
to be  compared  with  the persistence length of the dsDNA molecule $ A\simeq 50$~nm. This implies
that finite size effects may  be of some importance, specially the spatial obstruction 
 caused  by  magnetic tweezers.

The simplest way to implement  spatial constraints  is to introduce in the dsDNA 
elastic energy density  a  one-monomer potential $ V(\vr (s))$, where $ \vr(s)$ is the coordinate of the
monomer,  running along the chain of arc-length $s$.  
The ``worm-like-chain'' (WLC) model \cite{fixman,marsig,bouchiat99} describes rather accurately dsDNA elongation  
experiments. In its usual formulation,   
the sole  dynamical  variable  is the running tangent vector $ \vt(s)= \frac{d}{d \,s}\,\vr(s)$ and in that case, one  has to  write 
 $ \vr(s)=\int_0^s\vt(s') ds'$. The potential energy to be added to $E_{WLC}$  takes then an ugly non local
form $ \int_0^L ds\;  V \left( \int_0^s \vt(s') ds' \right)$.
   This difficulty can be  solved by formulating   the model in such a way that   $\vt(s)$ and $ \vr(s)$
behave as independent dynamical variables. 

 Numerous  authors addressed  this problem within the 
continuous version of the WLC  model. The  statistical properties of the molecular  chain   are obtained by 
solving a Quantum Mechanics problem, involving an  imaginary time $ -i \,s $. Using various arguments, 
they found that the Hamiltonian, allowing for spatial constraints,  is obtained  by adding  two  extra terms   
to the standard  WLC Hamiltonian  $ H_{WLC} =-\frac{1}{2 \,A }\, \nabla_{\vt}^2  - \vt \cdot \vf $.  
($ \vf $ is the stretching force  given in thermal units.)
The first is the so-called ``ballistic" term  $ \nabla \cdot \vt $  and the second  is the potential $ V(\vr)$   
\cite{Mag,Gom,Burk93,Burk97,Burk01,kier,saito,fried,Yama,helf,mors}. 
 This model has been applied to various problems in semi-flexible
polymer physics:
  the  flow of  semi-flexible polymers through cylindrical pores
\cite{Burk01},  the unbinding transition
 between semi-flexible polymers  attracted by directed polymers   \cite{kier}, the symmetric interface
 between two immiscible semi-flexible polymersÊ \cite{mors} and probably others...

In reference \cite {bouchiat06}, we have taken a different approach  by remaining  within
 a discrete version of the WLC model, which  has to be introduced  anyhow,
if one wants to write down explicitly the  functional integral   giving the  Boltzmann partition
function.  By using a simple trick,  we were able to write the partition 
function  as a multiple integral,  where the coordinates $ \vr_n$ 
and the tangent vectors $ \vt_n$  of the discrete molecular chain  are  treated as independent integration
variables. Using the transfer matrix formalism,  it is then possible to write
 down a recurrence relation between 
  adjacent intermediate partition functions  $ Z_n ( \vr_n, \vt_n ) $  relative to  chains having 
a  crystallographic length  smaller than the actual one:
\be
Z_{n+1} ( {\vr}_{n+1},{\vt}_{n+1})=\exp\((-b\,V({\vr}_{n+1})\)) \int\, d^2 \Omega( \vt_n)  
T_{WLC}(\vt_{n+1}\,  \vert\,  \vt_n)\,  Z_n( \vr_{n+1}-b\, {\vt}_{n+1},\vt_n )\,,
\label{recurZtrvf}
 \ee
where $T_{WLC}(\vt_{n+1}\,  \vert\,  \vt_n)$  is the transfer matrix relative to the unconstrained  WLC  model.
All the explicit computations performed in  reference \cite{bouchiat06} 
and in the present paper are based upon  the above iterative  construction, which  has    a suggestive  interpretation in terms  
of  a Markov random walk model in three dimensions.

The  confined dsDNA  configurations studied  in  reference \cite{bouchiat06}
were not fully realistic, since they do not account properly for the  spatial obstructions occurring  in magnetic  tweezers.
To  get a feeling about the orders of magnitude  involved  in  ``short" molecules, say with $ L\lesssim 10\,A$,   let us 
quote the results obtained in ref. \cite{bouchiat06}   for the   relative elongation upward shifts  induced by
   the spatial obstruction  of the  anchoring plate. ( By anchoring plate, 
we mean the plastic slab  upon which is stuck the initial molecular strand with the 
help of a ``biological glue''.)
   For a typical  stretching force $ F\simeq $ 0.3  pN, the  upward shift 
   is given by  $ 1.6 \, A/L $, which amounts  to 12 \%  for a 2 kbp molecule.
 In a magnetic tweezer, the free end of the dsDNA  molecule  is attached  to a magnetic spherical 
bead,  having a diameter of about one micron. In view of the above result, 
  the  bead  obstruction  effects  are certainly 
worth investigating. (To get  a very schematic  view of the various  constrained and unconstrained 
situations to be studied in the present work, see the upper graph of Fig.\ref{fig1} .)
Such a study will be particularly relevant for dsDNA molecules 
 with a few  kbp, when they are stretched by forces  within the range 0.1 to 0.5  pN. 
 
A theoretical analysis   of the spatial obstructions in a magnetic tweezer 
 is a difficult  task,  if one wants 
to treat it as a full three-dimension space problem, within the WLC  model.
The difficulties are about the same  in the approaches
based upon the solving of a  Schr\"odinger-like equation   or  
the transfer matrix iteration technique. The Monte-Carlo method, 
which could  perhaps be a  viable  approach, will not be considered in this paper 
because of  lack of competence of the author. We are going to  temper the  above pessimistic
views,   
by showing  that under well defined conditions the nucleotides do not really feel 
the curvature of the bead.  More precisely,  the bead surface can be reasonably  approximated by
its tangent plane at the molecule free-end anchoring point, assumed to be the lowest point of 
 the bead with respect to  the  force direction. It follows,  then,   that  under    realistic 
experimental conditions, the magnetic-tweezer obstruction can be simulated by  two repulsive  plates, 
normal to the stretching forces. There  is, however,  an important difference with the two-fixed-plate 
problem we have studied previously \cite{bouchiat06}. The initial molecular segment is still anchored to a fixed 
plate but the free end is now attached  to  a plate  which is no longer fixed, 
but in thermal equilibrium with the dsDNA molecule and the ambient fluid. We are 
back to a one-dimension space iteration  problem within transfer matrix technique.  It is
definitely more difficult than the two-fixed-plate problem but still manageable.

Volume-exclusion effects in tethered-molecule experiments have been studied recently by
Monte-Carlo techniques \cite{nelsonbeadexvol06}. The authors have considered the situation
where no stretching force are applied upon the bead. So, their significant work cannot be compared
with the present  one,  since  we  are dealing  with  stretched  molecules, having a relative  elongation larger
than $0.65$.  A finite-size effect analysis    appears also in connection with the 
 entropic elasticity of DNA molecule,  having  a permanent kink \cite{nelsonDNAkink}. 
In particular, the  authors  deal with the  boundary conditions to be satisfied by  the tangent vectors 
at the  two ends of the molecular chain. In the present work,   we have concentrated on the  spatial confinement,
imposed  upon the internal monomers, by the repulsive surfaces holding the
two free ends.    It is easily seen that  these  constraints insure  that
the  initial    and terminal  tangents vectors do satisfy automatically  the ``half-constrained" boundary conditions of 
ref.\cite{nelsonDNAkink}.

%%%%%%%%%%%%%%  SECTION II %%%%%%%%%%%%

\section{ A simplified model to describe the magnetic bead spatial  obstruction.}   
In this  section,  we would like to develop arguments to  justify  the replacement of  the bead surface 
by its tangent plane at the  anchoring  point,   assumed to be located 
at the bead  south pole  with respect to the force direction. The discussion will be performed 
  within a discrete  version of the WLC model. The molecular
chain  is    represented  by $ N $ elementary links, involving point-like ``effective" monomers, 
separated by a  length $b $,  much smaller than the
persistence  length   $A$. The    effective monomer  number $N$ is related to the contour length  $ L $  by the 
relation  $ N=L/b$. A  microscopic state of our model is then defined by the set of $ 2 N$   vector variables:
 $ \{ ( \vr_1,\vt_1)  \, ....  (\vr_n,\vt_n ), ...  (\vr_N,\vt_N ) \}$  with $ 1 \leq n \leq  N$, where
 $ \vr_n $  and $ \vt_n  $,    are respectively
 the  monomer coordinate  and   the unit tangent vector   such that   $ b\, \vt_ n=  \vr_ {n}-\vr_ {n-1}$.
In this paper the z axis  is parallel to the direction of the  stretching  force $\vF$ and has its origin at the  fixed-plate anchoring point.
One must stress  that the  variables $  \vr_N  $   and $  \vr_n$  with $ 1 \leq n \leq N-1$  have to  be treated on different footing:
$ \vr_N   $  is the   coordinate of the terminal monomer but it gives also the position of the bead.

 We shall, first, discuss the transverse fluctuations of the terminal 
monomer,  $  \langle x_N^2 \rangle $; they  are given in the thermodynamic limit $ L/A \gg 1 $  by a well 
known formula \cite{strick} :
\be
 \langle x_N^2 \rangle =\langle z_N \rangle \,(k_B \,T)/F =    \frac{\langle z_N \rangle}{  \alpha}\,A\  ,
\label{x2av}
\ee
where  $ \alpha $ is the dimensionless force parameter $ F \,A /(k_B \,T )$. This formula is used 
to calibrate  the force in magnetic  tweezer experiments and  was first obtained by a simple thermodynamic argument \cite{strick}.
In an  unpublished  note  \cite{stformCB},  we have given  a statistical mechanics evaluation of $ \langle x_N^2 \rangle$ within 
 the WLC  Model. We have recovered the formula (\ref{x2av}) in the limit $ L/A \gg 1 $ with  corrections  of the order  of 
 $ A/L $, which cannot be totally ignored for the  ``short" molecules considered in the present paper. For this reason,
 we have given an updated version of this note in the appendix. The  above value  of $ \langle x_N^2 \rangle$ will be used as a
benchmark in  our estimate of the transverse fluctuations of the internal monomers.  
To proceed it is convenient to introduce 
cylindrical coordinates for the effective monomer positions  $\vr_n= (z_n,\vrp_n)$ and the associated tangent vector
$ \vt_n =( \cos \theta_n, \vtp_n)$.  Our purpose is now  to estimate the thermal average of the square of the transverse distance
 between the n-monomer and terminal monomer: $ \langle \, ( \vrp_n -\vrp_N)^2 \rangle $. More precisely we are going 
to establish the inequality:
\be 
 \langle \, ( \vrp_n -\vrp_N)^2 \rangle < \langle\, \vrp_N^2 \rangle=
2 \langle\, x_N^2 \rangle= 2 \frac{\langle z_N \rangle}{ \alpha}\,A .
\ee

We have found convenient  here to work within  the WLC model in its simplest form, where 
only the tangent vectors $\vt_n= ( \cos \theta_n,  \cos \theta_n,\sin \theta_n \cos \phi_n , \sin \theta_n \, \sin \phi_n,   ) $
appear explicitly.  The n-monomer coordinate is then  given by the sum: $\vr_n= b\,\sum_{i=1}^{i=n} \vt_i $. 
The discrete  $WLC$  model  is  best formulated in terms of the transfer matrix connecting  two adjacent links: 
\bea
 T_{WLC} (\vt_{n+1},\vt_n) &
 &\propto \exp \((-\frac{A}{ 2 \,b} (\vt_{n+1}-\vt_n)^2  \))   \nonumber  \\
& & \propto  \exp \((\frac{A}{  b} ( \cos\theta_{n+1} \,\cos\theta_n+\sin\theta_{n+1} \,  \sin\theta_n  \cos ( \phi_{n+1}-\phi_n)  ) \)), 
\label{trmwlc}
\eea
where for simplicity we have omitted  the stretching energy term.
Our aim  is    to  prove that   the following difference is  positive:
\be 
\Delta_{\perp}= \langle\, \vrp_N^2 \rangle-\langle \, ( \vrp_n -\vrp_N)^2 \rangle= 
\langle\, \vrp_n^2 \rangle+ Ê2 \sum_{i=n+1}^{i=N}\sum_{j=1}^{j=n} \bra \vtp_i \cdot \vtp_j \ket.
\label{delperp}
\ee
  Clearly,  we have to show that the    thermal  average: $\bra \vtp_i \cdot \vtp_j \ket=
\bra   \sin\theta_i \,  \sin\theta_j  \cos ( \phi_i-\phi_j) \ket $ is positive.  Within the usual definition  of  the cylindrical
coordinates,  $\sin\theta_i \,  \sin\theta_j $  takes  only positive values while  $\cos ( \phi_i-\phi_j) $  can be  negative as well
as positive. To prove  the positivity of  $\bra \vtp_i \cdot \vtp_j \ket$, it is then  enough  
  to perform the thermal   average   over the azimuthal angle
$\phi_i$,  the longitudinal components  $ \cos \theta_i$  being frozen.
Assuming  that $ \phi_1$ is initially uniformly distributed, the overall  system is invariant upon
 any global rotation around the z-axis,  so that   we can replace the set of the $ N$  variables $\phi_i$
by the $ N-1$ variables $ \psi_i= \phi_i-\phi_{i-1} $ with $ 1< i \leq N$.  Using  the formula  (\ref{trmwlc})   
one gets  immediately the probability distribution of $ \psi_i$  :
\be
P_{i}( \psi_i) =   \exp \((\frac{A}{  b} \,  \lambda_i  \cos \psi_i \,\)) /  \(( 2 \pi I_0( \lambda_i) \)) \; , \; \lambda_i= \sin\theta_{i} \, 
\sin\theta_{i-1} >0 ,
\label{Ppsi} 
\ee 
 where, for convenience, we shall take $ -\pi \leq \psi_i \leq \pi$.
We have, now,   everything  we need to compute $ \bra  \cos ( \phi_i-\phi_j)  \ket $.  
 We first note that $\phi_i-\phi_j $ can be easily written as a sum of $ \psi_l$ angles: 
$$      \phi_i-\phi_j=  \phi_i-\phi_{i-1}+\phi_{i-1}- \phi_{i-2} ...\phi_{j+1}-  \phi_j = \sum_{l=j+1}^{l=i} \psi_l. $$
 The average $ \bra  \cos ( \phi_i-\phi_j)  \ket  $   is then  easily performed:
\bea  
\bra  \cos ( \phi_i-\phi_j)  \ket &=&  Re(   \prod_{l=j+1}^{l=i} \bra  \exp ( i \psi_l) \ket = \prod_{l=j+1}^{l=i}  \bra \cos \psi_l \ket , \\
\bra \cos \psi_l \ket & =& \int_{-\pi}^{-\pi} d \, \psi _l\, P_{l}( \psi_l) \, \cos \psi_l  .
 \eea
 By looking at the functional form of  $ P_{l}( \psi_l)$ in  eq.  (\ref{Ppsi})   it is easily seen that   $ \lambda_l  > 0$ 
implies $\bra \cos \psi_l   \ket  > 0$  and,  as a consequence,   the positivity  of $\bra  \cos ( \phi_i-\phi_j)  \ket $.
Since the final averaging will preserve this  positivity, we can conclude from eq. (\ref{delperp}) that $ \Delta_{\perp} >0,$  or
in a more concrete way: 
\be
 \langle \, ( \vrp_n -\vrp_N)^2 \rangle < \langle\, \vrp_N^2 \rangle=   \frac{ 2 \langle z_N \rangle}{ \alpha}\,A.
\ee
 Let us consider, now, an internal effective monomer going upward and crossing the tangent plane at a  transverse distance 
$ \vert ( \vrp_n -\vrp_N) \vert$ from the south pole. The maximum vertical distance $ \delta_{curv}$   which it
can travel  before hitting the bead surface is given by a simple geometrical argument: $  \delta_{curv}= ( \vrp_n -\vrp_N)^2 / (2 \,  R)$ 
where $ R$  is the  bead radius.
Performing the thermal averaging we get the  final inequality: 
\be
\langle  \delta_{curv} \rangle < \frac{ \langle\, \vrp_N^2 \rangle }{ 2 \, R} =  \frac{ \langle z_N \rangle}{R \,\alpha}\,A .
\label{dcurv}
\ee
The internal monomers  are subjected, in the vicinity of the bead,  to the chain tension  force which tends   to pull them  towards 
the anchoring point. As a consequence,  the above assumption  of a vertical  path  leads to  an overestimate of $ \delta_{curv}$, 
 so  relaxing this 
constraint   can only but strengthen the inequality of eq.(\ref{dcurv}).

 In order to quantify  the internal 
monomer ability to detect the curvature of the bead,   we are going to compare $ \delta_{curv}$
to other  experimental lengths. Let us begin by  the average elongation: $ \langle z(N) \rangle$.  We readily obtain  the 
 inequality: 
\be 
 \langle  \delta_{curv} \rangle   /   \langle z(N) \rangle  < \frac{ A }{ R \alpha } = 1.25 \times10^{-2} /R(\mu), 
\ee
where we have taken $ A= 50 \, nm $ and  $\alpha=4$, which corresponds  to a stretching force  $ F=0.31\, pN. $
$ R(\mu)$ stands for the bead radius given in micron.
A   more significant   comparison involves the  mean-square  longitudinal  fluctuation 
$ \Delta \, z_N= \sqrt{ \langle  z_N^2-  {\langle z_N\rangle} ^2\rangle }$. We have computed 
the ratio $  \Delta \, z_N / \langle z_N\rangle$ using a version of the WLC model
 incorporating the confining effect of  the fixed plate holding the initial strand \cite{bouchiat06}.
Taking   $\alpha=4$ and $ L=12\, A$, we  have obtained: $  \Delta \, z_N / \langle z_N\rangle= 0.083$.
Keeping the same value of $L$,  we  have derived    upper bounds  of  the  ratio $ \langle  \delta_{curv} \rangle   / \Delta \, z_N$,  for  
 $\alpha=2 \, ,4 , \, 5$ respectively:
\be
 \langle  \delta_{curv} \rangle   / \Delta \, z_N  < \frac{ \langle z(N) \rangle}{  \Delta \, z_N \ } \frac{ A }{ R \alpha }
=( 0.17 ,0.15, 0.13 )/  R(\mu). 
 \ee
  We conclude  that the  internal  monomer mean  free  path above the tangent plane  $\langle  \delta_{curv} \rangle  $
is  less   than    one sixth of the  mean-square  longitudinal    fluctuation of the bead  $   \Delta \, z_N$  when $ L=12 \, A$
and $\alpha \geq 2$.
  This  suggests   that,   under such conditions, the effective internal monomers  are not really able to   detect
the bead curvature  and gives a rather  strong justification for  
 the replacement of  the bead surface by its tangent plane at the anchoring point.
%%%%%%%  FIG0%%%%%%
\begin{figure}
\vspace{ 10mm}
\centerline{\epsfxsize=120mm\epsfbox{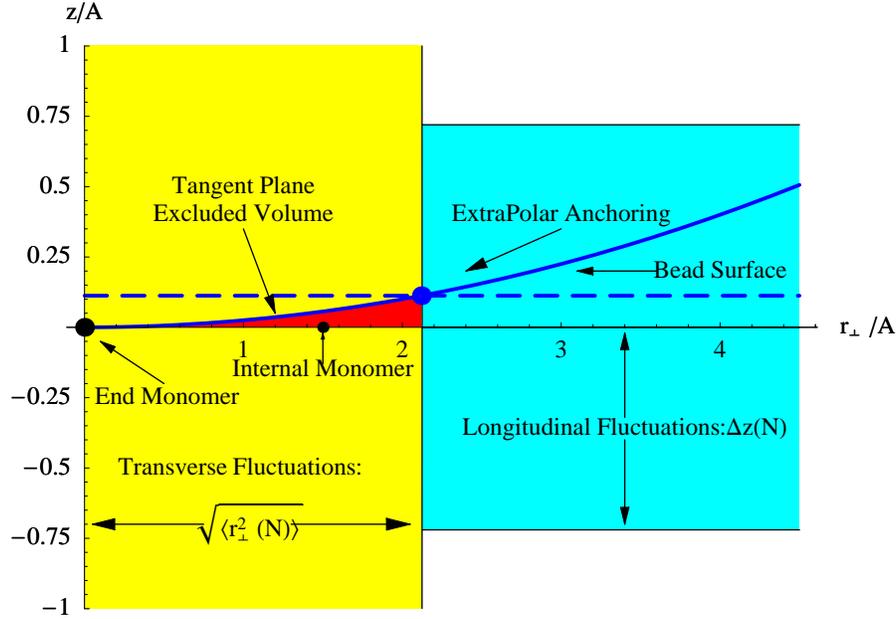}}
\caption{ \small  A schematic picture showing  the basic physical parameters 
 governing the motion of an  internal monomer
 (small  black dot) in the vicinity  of the  bead surface   tangent plane  at the terminal monomer ( big black dot)
anchoring point at the south pole. The figure represents   a section of the bead by a plane containing its center and  parallel to the
z-axis. The magnitude of the fluctuations are shown in the case of a force $F= 0.31$ pN and a contour length $ L=12\,A$. As indicated in
the text, the yellow  vertical  band gives an upper limit to the allowed transverse motion  of an internal  monomer with respect to the
terminal one. The red region   represents the  vertical section of the  excluded volume coming from the replacement of
the bead surface  (the blue circle arc) by its tangent plane. The emerald horizontal  band  gives the amplitude of the longitudinal
fluctuations  of the bead.  A comparison with the vertical width of  the red region suggests that the internal monomers will not be able to
``feel'' the bead curvature.  The big  blue dot represents  the case of an anchoring point lying away from the south pole. The plane
simulating the bead  is now the horizontal  plane $ z= z_N$ (its section appears as a blue dashed line). The excluded  volume is
subjected to  a  positive variation on the right-hand side  and a negative one on the other side; so there is clearly a compensation effect.
Furthermore, the selection procedure  described  in the text is affected by the transverse fluctuations of the bead. On this 
figure, we have assumed  for simplicity that the selection has been performed with $ F=0.31$ pN.  Taking instead 
$F= 2.7 pN$ would have reduced the selection angle $ \theta_{lim}$ by a factor 3. } 
\label{fig0}
\end{figure}

However,  one must keep in mind that the above conclusion hinges upon the simplifying assumption  that 
the end of the DNA molecule is stuck at the south pole of the bead. This condition will not be satisfied,   
unless some selection is  performed among the beads. This can be done in practice by slowly rotating the magnetic 
tweezer around its axis. The magnetic bead behaves as a compass and follows the rotating magnetic field.
If the experiment  is performed  at a high enough force $ F_{selec}$,
 the bead rotates around the vertical axis  passing  through the anchoring point  lying away from the south pole
( the big  blue dot appearing on Fig.\ref{fig0}).
As a consequence,  the center of the bead describes a circle of radius $ R \sin\theta_{an}$, where $ \theta_{an}$ 
is the ``latitude" of the anchoring point with respect to the bead south pole ({\it i.e} the angular distance between the black 
point  and the blue point on Fig.\ref{fig0}).
There is clearly a limitation in this bead selection procedure, coming from the transverse fluctuations of the bead.
 Only the beads satisfying the inequality
 $\sin {\theta_{an}}> \sin\theta_{lim} =\sqrt{ \langle\vrp_N^2\rangle \, ( F/F_{selec}) }/R $
 can be eliminated in practice. (We have made explicit the fact that the actual elongation experiment  is done at a force $F$
different from that used  in the selection procedure).
 Let us consider now a bead belonging to the selection. 
Replacing the tangent plane by the horizontal plane $  z=z_N $
induces a variation  (positive or negative ) of  the vertical free path of the internal monomers  of amplitude 
  smaller than   $  \sin\theta_{lim} \, \sqrt{ \langle\vrp_N^2\rangle  }= \sqrt{F/F_{selec}} \; \langle\vrp_N^2\rangle   /R    $. 
Taking into account the compensation apparent on Fig.\ref{fig0} and the eventual reduction factor $ \sqrt{F/F_{selec}}$
the average height of the excluded volume   can easily  be made 
 one order of magnitude   smaller than the longitudinal fluctuation of the bead $ \Delta z_N$.
 Our model is then expected  to be valid for the beads  selected  according to the above criterion.

%%%%%%%%%%%%%% SECTION III %%%%%%%%

\section{ A solution of the model and its physical interpretation.}
\subsection{ Computation Procedure.}
In this section, we are going to extend the  matrix transfer method of  reference \cite{bouchiat06} to the study of 
the statistical properties of a dsDNA molecule confined within  a magnetic tweezer. The two ends of the molecule are
respectively attached to a fixed plate and to a  magnetic bead immersed in a liquid which simulates the cellular medium.
The bead is subjected to an external  force normal to the anchoring plate. This pulling force is balanced by the tension
 of the  stretched molecule. As  shown in the above section,  the identification of the bead surface to its tangent plane 
  at the terminal monomer position is a fairly good approximation  if the anchoring point is close to the south pole.

 Our confinement configuration is  then defined by two parallel 
repulsive plates, the first - the anchoring plate - is fixed  and  the second - simulating the bead -  is in thermal equilibrium
with the attached molecule and the ambient fluid. 
With  proper  initial conditions,  the partition function of the system molecule-plus-bead
 is  invariant, first, under  rotations around  the stretching force direction and,  second,  under  translations
parallel to the plates. 
 We are going to use the same discrete version of the WLC model as in the previous section.
A  microscopic state of our model is then defined by the set of $ 2 N$ longitudinal variables: 
 \{$ ( z_1,\theta_1) \; .... ( z_n,\theta_n)\, ... ( z_N,\theta_N )$\} with $ 1 \leq n \leq  N$.

We proceed in two steps. We, first, assume that the   terminal monomer coordinate $ z_N $    has a  fixed value,
taken among a finite set chosen to be representative of the actual physical spectrum. The internal
monomer  coordinates  $  z_n$,  with $ 1 \leq n\leq N-1$,  are truly  stochastic variables,  associated with the partition function
$ { \cal Z}_n  ( z_n,\theta_n\vert  z_N ) $. For the moment,  $ z_N$  is treated as an external physical parameter. When
 $ 1 \leq n \leq  N-1 $,  the internal partition functions $ { \cal Z}_n  ( z_n,\theta_n\vert  z_N ) $  obey  a  recurrence relation 
which is just a rewriting of  a formula given  in Section 2.1 of reference \cite{bouchiat06}. The rule is  very simple:
 take the iteration equation   for the fixed-plate confinement problem and  identify   $ z_N$ 
 with   the  distance between the two  repulsive  plates $ L_0$:
 \bea 
{ \cal Z}_n  ( z_n,\theta_n\vert  z_N ) &=& \exp\((-b\,{\cal V}(z_{n}, z_N  )  \))
\int_0^{1} \hspace{-2mm}d\,(cos\theta_{n-1}) \, 
{\cal{T}}_{WLC}({\theta}_{n},{\theta }_{n-1}, f)\, \times \nonumber \\
& &{\cal Z}_{n-1}( z_{n}-b\,\cos\theta_{n},\theta_{n-1}\vert  z_N  ).  
\label{recurZint}
\eea
  The potential  ${\cal V}(z_n ,z_N  ) $, is written as the sum of two terms:
\be 
{\cal V}(z_N, z_n)=V_{pl} (z_n) +V_ {pl}( z_N- z_n),
\ee
 where the first one is associated  with the fixed anchoring plate and the second  with the fluctuating  plate simulating the 
magnetic bead surface.
The repulsive plate potential  $V_{pl} (z )$ is given in terms of the 
   rounded-off step function    \cite{bouchiat06}:
\be
\Theta( z,\Delta z)=\frac{1}{2} + \frac{1}{2} {\rm erf}( z/\Delta z) \,, 
\ee
where $ {\rm erf}(x) $  is the  ``error" function : $ \frac{2}{\sqrt{\pi }} {\int}_0^x \exp (- \,t^2) \,d\,t $ and $\Delta z$
a smoothing length  assumed to be $\sim \,A $.  
For the sake of simplicity, it is convenient to give directly the associated Boltzmann factor:
\be
 \exp\((-b\, V_{pl} (z )  \))    =\Theta( z,\Delta z) .
\ee
 The conditional probability distribution  ${\cal{T}}_{WLC}({\theta}_{n},{\theta }_{n-1}, f)$ is obtained 
by performing   an  azimuthal average   of  the transfer  matrix given by equation  (\ref{trmwlc}):  
 \bea
{\cal{T}}_{WLC}({\theta }_1,{\theta }_2, f) & = &\exp-\lbrace \frac{A}{b}\, \left( 1  -
                                  \cos {\theta }_1\, \cos{\theta }_2  \right) 
   +\frac{b\,f}{2} \left( \cos {\theta }_1 + \cos {\theta }_2 \right)  \rbrace \nonumber\\
& & \times  \,I_0 ( \frac{ A \sin{\theta }_1\, \sin{\theta }_2 }{b}   )
\label{TWLC }\, ,
\eea   
where $f$  is related  to the stretching force  by $ F = f \,   k_B \, T.$
 We  compute the    partition function  relative to  the last internal  monomer, for a fixed terminal monomer, 
${ \cal Z}_{N-1}  ( z_{N-1},\theta_n\vert  z_N ) $, 
 by running, up to $n=N-1$, the iteration  process defined by  equation (\ref{recurZint}). To perform   the relevant 
recurrence process we have used {\it Mathematica}  codes where analytical and numerical computations are 
intertwined.  For more details about our procedure, see  the SectionV of  reference \cite{BouMez00}.

   The second  step  ( a short one !) is to  compute the partition  function $ Z_N(  z_N, \theta_N ) $ 
 relative to the terminal monomer.
It is   easily obtained  from the formula:
\bea 
 Z_N  ( z_N,\theta_N ) &=& \exp\((-b\,V_{pl}(z_{N})  \))
\int_0^{1} \hspace{-2mm}d\,(cos\theta_{N-1}) \, 
{\cal{T}}_{WLC}({\theta}_{N},{\theta }_{N-1}, f)\, \times \nonumber \\
& &{\cal Z}_{N-1}( z_{N}-b\,\cos\theta_{N},\theta_{N-1}\vert  z_{N}  )  
\label{Zterm}.
\eea
Finally, the probability  distribution relative to the 
terminal monomer  longitudinal coordinate  
- it is also the longitudinal  distance of the bead  south pole  from the fixed  plate - 
is given as follows: 
\bea 
P_{N}(z_N)&= &\frac{1}{{\cal N} } \int_0^{1} \hspace{-2mm}d\,(cos\theta_{N}) \,  Z_N  ( z_N,\theta_N ),  \nonumber \\
{\cal N}&=&\int_0^{L} \hspace{-2mm}d\,(z_{N}) \, \int_0^{1} \hspace{-2mm}d\,(cos\theta_{N}) \,  Z_N  ( z_N,\theta_N ).
\eea
\subsection{Result and Physical Interpretation.}

\begin{figure}
\vspace{ 10mm}
\centerline{\epsfxsize=120mm\epsfbox{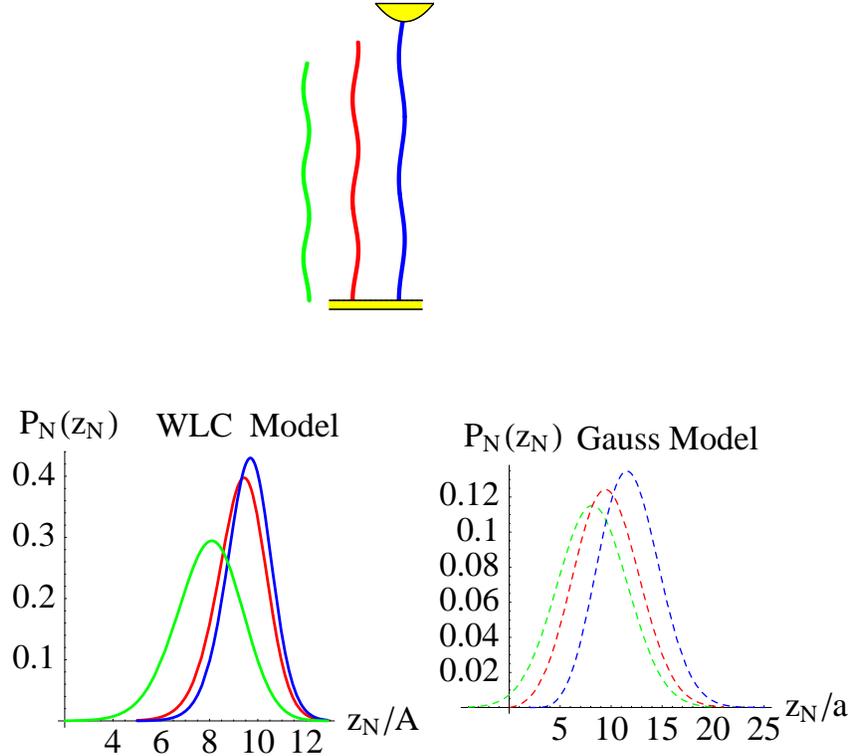}}
\caption{ \small  Probability distributions  of the terminal coordinate $ z_N$,  relative  to
 different spatial obstruction conditions, as predicted by two models. The continuous curves appearing
on the left-hand lower  graph  have been obtained  within the  WLC  model, 
while the dashed ones   on the right-hand graph  are relative to the
Gaussian model.  As indicated in the picture  given in the upper graph, the green color
 corresponds to a molecular chain free of any external spatial constraints.
The  red and blue colors  are associated  respectively 
with   two  kinds of spatial obstruction:  restriction to the  upper half space by  an anchoring fixed 
plate and confinement between the  fixed plate and the  fluctuating bead,  holding the terminal monomer.  } 
\label{fig1}
\end{figure}

The probability distribution  $ P_{N}(z_N) $ is displayed  upon the left  graph of  Fig.\ref{fig1} as a continuous  blue curve. 
It appears together with green and  red curves,  corresponding,  respectively, to the two  situations: absence of constraints  and  
presence of a fixed anchoring plate.
 Two   features are conspicuous: 
\begin{itemize}
\item   The red and blue curves  are very close,  while the green and red  ones are much further apart, showing that the obstruction 
effect of the bead is much smaller than that coming from the fixed anchoring plate.
\item  Perhaps even  more surprising,   
the  bead  obstruction  effect  reduces to  a slight  push  upwards  given  to 
the terminal monomer,  while one might have expected naively the reverse.
\end{itemize}

  In the two-fixed-plate configuration studied in  our previous work \cite{bouchiat06},
  the confinement effects  upon  $ P_{N}(z_N) $ were very  spectacular  ( see  Figure 4. of this reference).
   It must be  stressed  that the physics involved was very different from that prevailing  in  the present paper.
 Indeed,  the terminal monomer, supposed to be  attached to a nano-magnet,  was free to move between two
diamagnetic fixed  plates.  When the elongation in absence of plates  was larger than the two-plate distance, the probability  
distribution  $ P_{N}(z_N) $
 was flattened against the upper plate  by the stretching  force. { \it  No such phenomenon appears here
since the  terminal monomer is stuck upon the plate simulating the magnetic bead. 
 Only the internal  monomers  feel the repulsion of the 
magnetic bead and it is through their intermediary that the terminal  monomer is affected by the spatial obstruction of the bead. }

%%%%%%%  FIG1%%%%%%

A plausible mechanism goes as follows: the  internal monomers 
 are  pulled upwards by the  string  tension  force.   When they collide with the bead,   they are expected to give 
  a small  upward  push to the bead surface. In other words, they are exerting  
an upward   pressure on  the bead,  which is transmitted  to the terminal monomer, leading  to a small 
increase of the stretching  force.

The positivity of  the variation of the elongation $\delta  \bra z_N \ket  $
can  also be obtained  by a simple  thermodynamic argument, valid in the limit $\delta  \bra z_N \ket / \bra z_N \ket \ll 1.  $
The  internal energy of  the  bead-plus-DNA system  receives  a  positive contribution   $ \delta U $  
    coming from  the  repulsive interactions:  $ \sum_{n=1}^{N-1}  V_ {pl}( z_N- z_n) >0$. Writing that the
variation  of the free energy ${ \cal F}= U-  \bra z_N \ket \, F $ vanishes
 at equilibrium, one gets: $ \delta  \bra z_N \ket \simeq \delta U/F >0 $.

In order to get a confirmation of the above picture,  we have repeated our  computations
 within  the soluble ``Gaussian  Model'',  often used to describe  ``flexible'' polymers. It involves 
a chain of point-like monomers connected by harmonic springs. The continuous limit of the chain 
is  a  string described by the  elastic energy linear density:
 ${\cal E}^{gaus}(s)=  \frac{ 1}{2\,a}\,(\dot \vr(s) )^2+ V(\vr)$, where  
$\dot \vr(s)$ is the derivative  of the monomer coordinate $ \vr(s)$ with respect to the string arc-length $s$ and 
  $ a^{-1} $  is   proportional to the rigidity of the spring connecting   nearest-neighbor monomers. The value chosen  for the 
parameter $ a $  guarantees that the relative elongation  $ a f $, in  absence of  spatial constraints,  is  the same as in the WLC model.
  The  monomer number $n$ is related to the  arc-length $s$   by $s=n b $ with   $b/a \ll 1$.
If we identify  $  V(\vr)$   with the potential $ {\cal V}(z_n ,z_N  ) $   introduced above, 
the internal monomer  probability distribution, $ P_{n}^{int}(x_n,y_n, z_n)$, 
 factorizes into three independent distributions relative to  each component.  
In the continuous limit, the statistical  properties of the internal monomers 
are   easily obtained by exploiting the analogy with  a QM
problem  involving    the following   simple Hamiltonian:
\be \widehat{H}_{int}  = -\frac{a}{2}\,\frac{\partial^2 }{\partial z_n ^2}+{\cal V}(z_n ,z_N  ).   \ee
The Hamiltonian relative to  the  terminal monomer $ \widehat{H}_{term} $  is  obtained by performing  the replacement : 
 $ z_n \rightarrow  z_N  \; ,  \;{\cal V}(z_n ,z_N  )   \rightarrow V_{pl}( z_N) $. The probability distribution 
for the terminal monomer   involving a fixed anchoring plate  together with  a pulling one,  in  thermal equilibrium with the "Gaussian``
polymer chain,  reads then  as follows:   
\be
P_{N}(z_N)=\int dz_{N-1} \,  \exp(f \,z_N)\bra z_N \vert \exp\(( -b\ \widehat{H}_{term}\))\vert z_{N-1} \ket
 \bra z_{N-1} \vert \exp( -b \,(N-1) \,\widehat{H}_{int})\vert z_0 \ket . 
\label{Pzterm}
\ee
The results obtained      within  the ``Gaussian  Model" are
 displayed on the right-hand  graph of Fig.\ref{fig1}.  The  obstruction effects  are qualitatively similar 
   to those obtained within the WLC model but   the  chain-elongation increase  induced by ``the bead''
$ \delta  \,\langle z_N \rangle $  is much larger. This amplification  can  be understood  by  noting  that
the 	 ``Gaussian  flexible'' monomers, being allowed to  move much more freely, have a larger collision rate with ``the bead''.
%%%%%%%  FIG2%%%%%%
\begin{figure}
\vspace{ 10mm}
\centerline{\epsfxsize=120mm\epsfbox{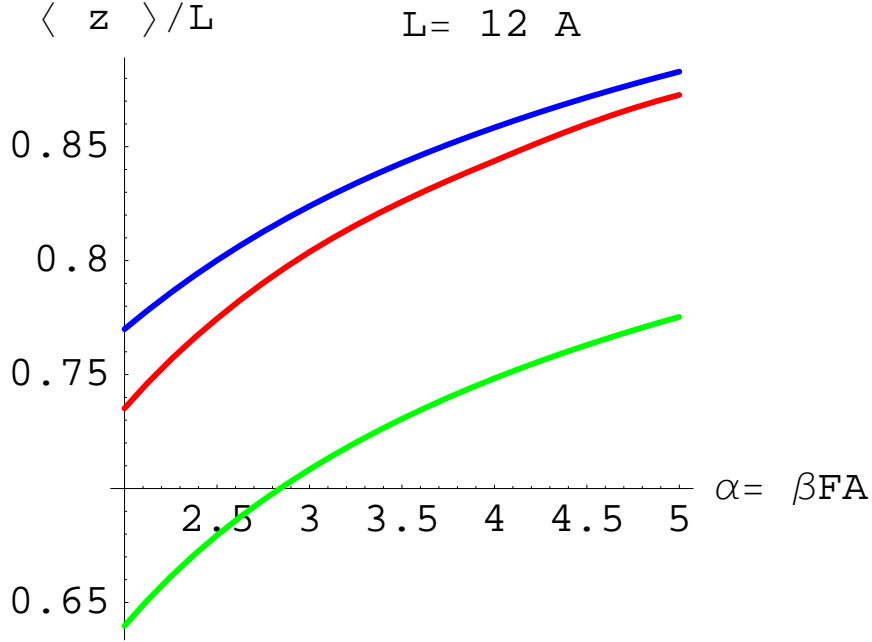}}
\caption{ \small    This figure    displays three curves,  giving the ``short" dsDNA  relative elongation versus 
the stretching force, corresponding to different spatial constraints. 
   Going from top to  bottom,
the blue  curve is the prediction of the WLC model, incorporating both the bead and the anchoring plate spatial
obstruction. The red curve accounts  for the sole   effect of the anchoring plate. The  green curve,   lying
significantly below,  has  been  computed  in absence of any spatial constraints. These curves emphasize
the dominant role played by  the anchoring plate.}  
\label{fig2}
\end{figure}

We have studied the magnetic tweezer spatial constraint effects on the dsDNA   elongation within the stretching force range
defined by $ 2 \leq \alpha \leq 5 $, in the case of  a contour length $ L= 12 \,A$.  The elongation curves are displayed 
in Fig \ref{fig2} for three configurations :   no  space constraints  ( green line),  an anchoring plate barrier ( red line) 
and an anchoring plate barrier together with the magnetic bead ( blue line ). As discussed above,  
spatial obstruction effects from the bead,  in  thermal equilibrium, 
 lead   to a small   increase  - about few $\%$- of  the elongation. 
 It is to be compared   with the four times larger effect induced   by the anchoring  fixed plate, where the repulsive character  of the
fixed barrier is  clearly  exhibited.

%%%%%%% SECTION IV %%%%%%%%

\section{ The non-extensive behavior of the  DNA  elongation   in  presence of  spatial  constraints.}

In this section we shall study, for fixed forces, the variation of the elongation $ \langle  z(N) \rangle  $ with respect 
to the contour  length $L$. This analysis has been  performed previously in ref. \cite{bouchiat06}. In the case of   
fixed  plates, it was found that $\langle z(N)\rangle $  is no longer an extensive variable with respect to $L$. 
The effects were very  spectacular  in  the two plates configuration
when the distance $ L_0$ between the two plates is smaller than  the elongation in absence of spatial constraints. 
Conversely, when  $ L_0 > L$,  the molecule feels only the anchoring plate and the extensive behavior of 
 $ \langle  z(N) \rangle  $ is perturbed in a very simple way. For stretching forces such that $ \alpha \geq  2$ the 
derivative of elongation $d/dL\,\langle  z(N) \rangle$ does not vary with $L$ and stays very close to the {\it constant }
  $ {\langle  z(N) /L\rangle}_{WLC}  $  predicted,  for a given force,   by the WLC model  
in absence of spatial constraints. The only modification is the apparition
in the  elongation of an offset term independent of $L$. In more precise words,  the elongation $ \langle  z(N) \rangle  $
can be written as follows  when $ L > 2  \, A$:
\be
\langle  z(N) \rangle= L\,{\langle  z(N) /L\rangle}_{WLC}\,(1+\epsilon) + A\,\Delta _{ofs}( \alpha),
\label{linearfit}
\ee
where $ \vert \epsilon\vert < 10^{-2} $ and the dimensionless offset $\Delta _{ofs}( \alpha)$ is  a slowly decreasing function
of  $ \alpha $, which takes values of the order of unity when $ 2 \leq \alpha \leq 5 $.
%%%%%%%  FIG3%%%%%%
\begin{figure}
\vspace{ -10mm}
\centerline{\epsfxsize=100mm\epsfbox{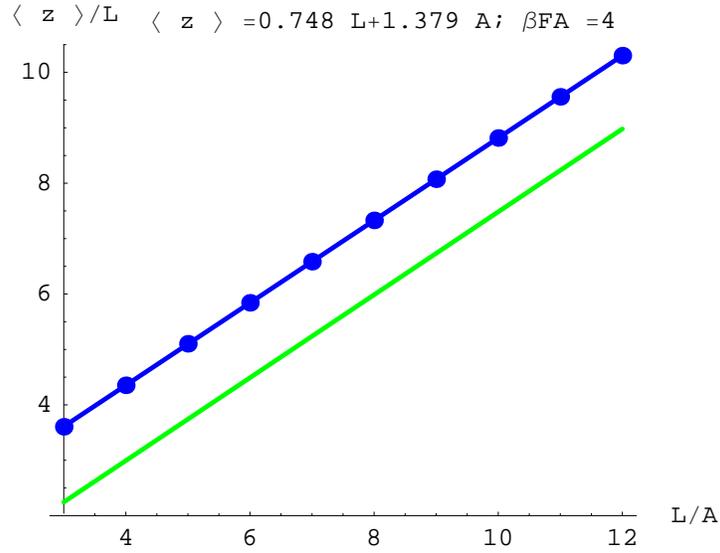}}
%\centerline{\epsfxsize=100mm\epsfbox{espsi}}
\caption{ \small  dsDNA elongation versus 
   contour length,  within the interval 
 $ 2\, A \leq L \leq  12\, A$.  The blue points 
have been obtained  from a version  of the WLC model, implementing  
 the  magnetic tweezer spatial  obstructions. They are well fitted by a straight line (the blue line), having a slope 
very close to the prediction of the unconstrained  WLC model, given by the green  line. 
It is to be noted  that the extrapolated fitted  line  does not pass through  the origin, as it should 
   if the corrected elongation were still an extensive quantity. The constant offset  term, 
responsible for   this non-extensive behavior,  gives rise  to a $15\%$ upward shift  of the elongation when  $ L=12 \,A$.}
\label{fig3}
\end{figure}

In the present paper, we have performed  - within our model - the same analysis in presence of the magnetic bead in thermal   
equilibrium  with the dsDNA. We have also  found  a   linear variation of the elongation with respect to $L$, similar  
to that given by  equation (\ref{linearfit}); the only difference is an increase  of the offset  function   $\Delta _{ofs}( \alpha)$
by a few tens of percents. As an illustration, we have plotted on  Fig. \ref{fig3} the result of a linear fit   (the blue line) involving    the
 elongations $\langle  z(N) \rangle$ relative to a given set of contour lengths.
They were   computed with the method described in section II  for $ \alpha= 4 $, when $L=N \,b  $ 
takes  10 equally spaced values within the interval    $ 2  \,A \leq L \leq 12 \,A $. The slope coming from the fit coincides, 
 to better than  1\%, with the prediction  of the unconstrained WLC model, which  appears as a green  line on Fig. \ref{fig3}.
 Similar  features  hold true for the  linear  elongation  fits  performed on the  results obtained   with     seven 
 equally spaced  values of $\alpha $ within the  interval $  2 \leq \alpha \leq 5$. 
Although we have verified the validity of the linear fit  for values of $L >12 A$  and  $ \alpha  \geq  2$ , it is certainly
not valid for  $  \alpha  \ll  2$.  We have, indeed,  found   in ref. \cite{bouchiat06} (section 2.2)
 that the elongation upward shift contains an extra term $ \propto   \sqrt{L}$  when $ 0 \leq \alpha \leq  1$, with a coefficient
which exhibits a very steep decrease with $ \alpha $.

%%%%%%%  FIG4%%%%%%
\begin{figure}
\vspace{ 10mm}
\centerline{\epsfxsize=100mm\epsfbox{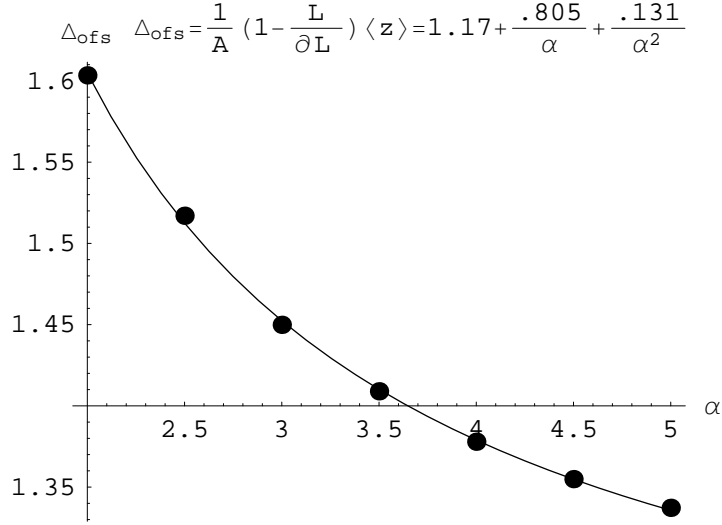}}
\caption{ \small    Numerical values of the offset term $\Delta _{ofs}( \alpha) =      \(( 1  -L \, / d\,L \))\bra z(L) \ket /A  $  
versus seven representative values of $ \alpha$. A second order polynomial involving $ \alpha^{-1} $ 
provides a rather accurate interpolation within the interval $ 2 \leq \alpha \leq 6 $. 
}
\label{fig4}
\end{figure}
 We have plotted on Fig.\ref{fig4} the 
values  of the offset function    $\Delta _{ofs}( \alpha)$    coming from the corresponding elongation linear fits. For the sake
of convenience, we have  performed a third order polynomial fit, using   $ \alpha^{-1}$ as variable:
\be
\Delta _{ofs}( \alpha)=1.6942 + 0.8052 \, \alpha^{-1} + 0.131213  \, \alpha^{-2}. 
\label{offset}
\ee
 No particular physical 
significance  is to be attributed to the functional form chosen for the fit other than the fact it gives a rather simple
 and accurate interpolation between the calculated values. 
The formulae (\ref{linearfit})  and (\ref{offset}), together with the accurate values  for
 $ {\langle  z(N) /L\rangle}_{WLC}$  given in  reference \cite{bouchiat99},
lead to a one \%  evaluation of  the dsDNA  elongation, corrected  for the spatial
constraints induced both  by the anchoring plate and the magnetic bead. A word of caution:
    formula (\ref{offset}) is not to be trusted   if it is used outside the range   $  2 \leq \alpha \leq 5$  and for $L< 2 \, A$;
in particular, it does not give the correct limit for $ \alpha \gg 1 $ since one expects, on physical ground,  that 
$ \lim _{\alpha \rightarrow \infty} \, \Delta _{ofs}( \alpha)=0$.
 %%%%%%%%%%% SECTION V%%%%%%%%%%%
\section{The empirical determination of the persistence length for spatially confined ``short" dsDNA molecules.}

One may consider that  the modification of the dsDNA elongation induced by the magnetic tweezer obstruction 
effects is, after all, not  so dramatic  since for the "typical`` case $ \alpha =4 \; \text{and}\; A/L = 1/12$,  
it amounts to an increase of  $ 15\, \%$. However, if the elongation together with transverse 
fluctuation measurements are used  to get a determination of the persistence length,  the situation becomes 
much more serious. We would like to show  that  the raw data of $ \bra z(L) \ket $,
without a subtraction  of the offset correction $\Delta _{ofs}( \alpha) A/L$,  could lead to totally unreliable 
values  of $ A$.  The  basic  principle behind this determination  of $ A$ is to combine two  formulae giving the 
stretching force $ F$ in terms of     $  \bra z(L) \ket\; \text{and}\;\bra x^2 (L) \ket$: the first one  is 
 the Strick {\it et al.} \cite{strick}  formula,   which is discussed in details in the appendix; 
 the second one  is the relation between  the reduced force parameter $ \alpha $  and the relative elongation 
  $ u  =  \bra z(L)\ket/L $.     In the case of the unconstrained  WLC  model, it takes
a  scale invariant  form: $  \alpha =  {\cal F }(u) $, where $ {\cal F }(u) $ is  a numerical function,  which 
interpolates accurate numerical results \cite{WLCelong}. By a simple algebraic manipulation,  one 
arrives at the relevant formula, valid only within the unconstrained WLC model:
\be
A= \frac{ \bra x^2 (L) \ket}{ \bra z(L) \ket} \,   {\cal F }(\frac{\bra z(L) \ket}{L} ).
\label{Avselong}
\ee
It cannot be applied bluntly to the case of ``short"  DNA molecules studied  in the present paper.
Let us call $ u_{raw}=   \bra z(L) \ket_{raw}/L $ the relative elongation involving   the non-subtracted elongation
    $\bra z(L) \ket_{raw}$, given in our model 
 by  equation (\ref{linearfit}) and $ A_{raw}$ the corresponding persistence length,   
obtained by plugging   $u_{raw}$ in 
equation  (\ref{Avselong}).  Ignoring the effect of spatial constraints upon the 
transverse fluctuations  $\bra x^2 (L) \ket$, one arrives immediately to the ratio: 
\be
A_{raw}/A= \frac{  u \,{\cal F }(u_{raw})}{ u_{raw}\, {\cal F }(u) }.
\ee
  Performing the numerical evaluation   for the ``typical" case $ \alpha =4 \; \text{and}\; A/L = 1/12$,   
one gets the rather spectacular result: $ A_{raw}/A =2.9  $. Such a  large number  cannot be
be explained by the finite size correction to  the Strick {\it et al.}  formula,  estimated in the appendix.
It is coming from the fact that ${\cal F }(u)$  exhibits a very sharp increase  when $ u \geq 0.75 $, due to 
the presence of a pole singularity at $ u=1.$ (See the foot note \cite{WLCelong}.)
If one subtracts from $ u_{raw} $  the offset correction $\Delta _{ofs}( \alpha)A/L$,  
the  ratio gets back to a value very close to 1. 

If the non-extensive behavior of the elongation has the simple  linear behavior  described in the previous 
section, the required subtraction may be achieved empirically.   The procedure  involves
 elongation measurements  upon  three  (or more !) ``short" molecules with  different  contour lengths, 
say  $ L_1=  5\, A \, , \, L_2=10 \, A \; \text{and} \; L_3= 15 \,A $,  subjected to the same stretching force. 
(This latter requirement  may be difficult to satisfy with precision.)
A linear fit  to the  data  will lead to an empirical determination of $ A\,\Delta _{ofs}( \alpha)$,  
 together with a  verification of  the near  equality  between the fitted slope  and the prediction  of the unconstrained  WLC  model.
If the result of the  latter test is positive, then one can proceed to the required subtraction from the elongation and plug  the result
in equation  (\ref{Avselong}). One  should get in this way  a value of $A $  close to that obtained  for ``long" molecules, 
say with $ L/A \geq 100$.  Finally, the comparison of the empirical 
value of $\Delta _{ofs}( \alpha)$ with the prediction given in the previous
section will provide a further significant   test of our model for the spatial obstruction effects in a magnetic 
tweezer.

%%%%%%%% CONCLUSION%%%%%%%%%%

\section{ Summary and perspectives}
 In the present paper, we have proposed a theoretical model for the spatial-constraint 
corrections   to  the dsDNA elasticity  measured with a magnetic tweezer.
 An evaluation  of the obstruction effects of the fixed anchoring plate had been given 
previously \cite{bouchiat06} and here  we  have concentrated  upon the magnetic bead which is
 attached to the molecule free end. The main and somewhat unexpected result of this work 
is that the magnetic bead obstruction effects give rise to a slight  upward shift of the elongation,
about four times smaller than  the anchoring plate effect.

\subsection{ Synopsis of the paper.}
\begin{itemize}
 \item In section II,   we have developed  theoretical arguments  to justify the replacement of the bead surface 
by  its tangent plane at the anchoring point, assumed to be located  close to the  bead south pole,
defined by   the force direction. As a first step, we prove, within a discrete  version of the WLC  
model,   that the mean square transverse distance between  an arbitrary internal ``effective'' monomer and 
the anchoring point is smaller than the transverse fluctuations $ \bra  x_N^2 \ket$  of the terminal  monomer. This 
latter quantity is given by the Strick {\it et al.}  formula \cite{strick}  in terms of the force and the molecular elongation. Then we 
proceed, by a simple geometrical argument, to the derivation  of a lower  bound  for the internal monomer
mean free path $ \delta_{curv}$ above the tangent plane. For ``short" molecules ( $L \lesssim 10 \, A $),  stretched by a
force $ F \geq 0.3$ pN,  $ \delta_{curv}$ is about six times smaller than 
the root mean square longitudinal fluctuations of the terminal monomer, if the bead radius is larger than one micron.
This result suggests strongly that the internal monomers do not  really ``feel" the bead curvature. 

 \item In section III,  we have given a transfer matrix solution  of our confinement model,  involving a discrete chain of 
   $N$ effective monomers  with the two extremities attached to  a pair  of  parallel  repulsive plates. The initial monomer is 
anchored upon a fixed plate,  while the terminal one is  stuck to a fluctuating plate, in thermal equilibrium with the attached
chain and the ambient fluid. In the first  step, the terminal monomer longitudinal coordinate has a  fixed  
value $ z_N $, taken among a representative set. The partition function of   the chain of  $ N-1$
 fluctuating internal monomers    are then   obtained
by solving,  for each fixed value of   $ z_N$,  a two-fixed-plate confining model, using the method of  ref. \cite{bouchiat06}. 
In the second step, we obtain the  partition function of the terminal monomer by applying to the last internal monomer  
the transfer matrix relative to the spatially unconstrained  WLC model. The probability distribution of the terminal 
monomer distribution $ P_N( Z_N)$ exhibits two remarkable features, when it is compared to the same distribution in
absence of  magnetic bead obstruction.  First, the two curves,  appearing in Fig. 1  as blue and red continuous lines, are very close.
 Second, the effect of  the bead reduces  to a slight upward push given to the terminal monomer.
One may have expected naively exactly the reverse: a significant downward push  from the ``bead". 
However, in contrast with the two-fixed-plate problem \cite{bouchiat06}, only the internal monomers 
feel  directly the repulsion from the ``bead", while the terminal monomer stuck on the bead  don't.  So, one has to invoke an indirect 
effect involving the internal monomers.  A plausible candidate is  the upward pressure they exert on the bead, 
which is transmitted to the terminal monomer. As long as we are dealing with a small effect, 
a simple thermodynamic argument leads also to
an upward shift  of   the elongation. We have obtained 
a qualitative confirmation  of the whole picture  within the soluble ``Gaussian" model, used   for flexible polymers. We have  found 
   similar effects, but about seven times larger  in typical cases. This  amplification reflects the fact that the internal
 monomers of a flexible polymer, being allowed to move more freely,  have larger colliding rates with the ``bead". 

\item The section  IV is devoted to an analysis    of the non-extensive behavior of the DNA elongation induced 
by the magnetic tweezer confinement effects. We have, indeed, found that in DNA molecules,  having a contour length 
$ L \geq 2\,A$,  the elongation $ \bra z(N) \ket$  is no longer an extensive quantity, within the   force range $ 2 \leq   \alpha \leq  5$.
The derivative of the elongation with respect to L, $ d /d\,L   \bra z(N)  \ket $, stays 
very close to the constant value predicted by the unconstrained $ WLC$  model.
 The sole  non-extensive effect is the apparition in   $ \bra z(N) \ket$ of an offset term $ \Delta _{ofs}( \alpha) \, A $
independent of $ L $. In  other words, the  elongation-versus-$ L$  curve  is still a straight line  but it 
does not go through the origin. The dimensionless offset term $ \Delta _{ofs}( \alpha) $ decreases slowly from 1.6 
to 1.3 within the interval $ 2 \leq  \alpha \leq  5$  and it is well represented by  a second order polynomial in $ \alpha^{- 1}$. 
For the ``typical'' case $ \alpha =4 \ \rightarrow F= 0.3 \, \text{pN} \; \text{and}\; A/L = 1/12$, the non-extensive  correction
to  $\bra z(N) \ket/N $  amounts to $ 15 \%$. 

\item In the  final  section,  we investigate the possible influence of  the magnetic tweezer  confinement effects
 upon  the determination of the persistence length $A$,  from elongation and transverse  fluctuations
measurements. 
 We consider the case of ``short" molecules of about 2  kbp, stretched by a force $ F \simeq 0.3$ pN. 
  Plugging in the basic formula  a ``raw" relative elongation, uncorrected for  non-extensive effects,
  leads to an  overestimate of $A$  by a factor 3  with respect to ``long"  molecule values. We suggest an 
empirical way to perform the required  subtraction from  the measured  elongation. 
\end{itemize}
\subsection{ Possible extension of the present work to super-coiled dsDNA molecules.}
%%%%%%%%%% FIG 5 %%%%%%%%%%%%%%%
\begin{figure}
%\vspace{ 10mm}
\centerline{\epsfxsize=100mm\epsfbox{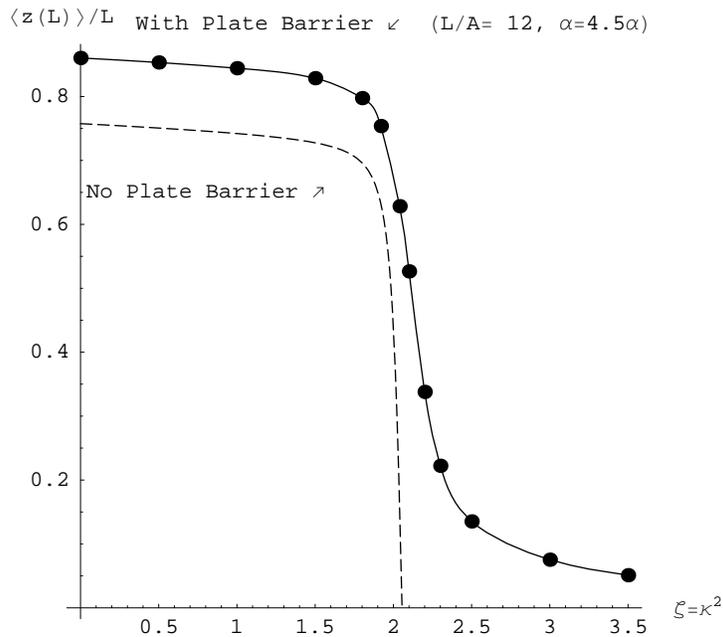}}
%\centerline{\epsfxsize=100mm\epsfbox{espsi}}
\caption{ \small  Two  ``hat" curves, giving the elongation of a super-coiled molecule versus the torque.
The upper curve  incorporates the obstruction effects of the anchoring plate, which are  ignored in the 
lower one.  } 
\label{fig5}
\end{figure}
	
 In references  \cite{BouMez98,moroz,BouMez00} the WLC  model has been generalized to a 
Rod Like Chain  (RLC) Model, involving both bending and twisting rigidities.
This makes possible the study    of   super-coiled dsDNA   entropic elasticity
below the denaturation threshold.

 One can readily modify  the RLC  model in order to  incorporate spatial constraints.  The  recurrence
relation for  the  partition function $Z_n( z_n,\theta_n,\kappa) $   relative to a super-coiled  DNA molecule, with a  given torque  
$ \Gamma=k_B \,T \kappa $  acting upon  its free end, is obtained by 
 performing  in the r.h.s. of the recurrence relation (\ref{recurZint}) the replacement:
 $ {\cal{T}}_{WLC}(\theta_{n+1},\theta_n,f) \rightarrow  {\cal{T}}_{RLC}( \theta_{n+1},\theta_n,f,-\kappa^2) $, where 
$ {\cal{T}}_{RLC}$ is given explicitly in ref.\cite{BouMez00}. 
The anchoring-plate barrier is  expected to  
have   significant  effects upon the so-called ``hat curves", giving, for a fixed force,  the relative elongation 
versus the super-coiling reduced  parameter $\sigma$. Let us take  the  ``low" force case  studied in ref.   \cite{BouMez98} ,  $
F\simeq 0.1$~pN, where the  RLC model  ``hat curve'' dips steeply into the negative $z$ region when  $ \vert \sigma \vert \geq
0.03$.  This effect is  attributed to the creation of plectonem structures which  are allowed to wander in the  $ z <0$ half space,
because of the   vanishing of their  stretching-potential energy. Therefore, 
we can expect important modifications  of  the torsion elasticity  if the spatial constraints are incorporated in the RLC model.

This effect  is illustrated in Fig. 5,  which displays 
the results of  a preliminary computation giving the relative elongation $ \bra Z(L) \ket $  versus 
the torque $ \Gamma= \kappa \,k_B \,T $ for a ``short" DNA molecule ( L/A=12)  and $F=0.33 \,pN$.
The solid line is a spline fit connecting the  points  obtained with the RLC model, taking 
into account the anchoring plate obstruction effects. The dashed  line is relative to the RLC model free of
any spatial constraints. The dip of the  latter curve  into the $ z  \leq 0$  half-space   reflects 
the formation of plectonem-like configurations. For $  0< \kappa<1$,  the plate barrier effect reduces  to an upward  shift of 
the elongation,  similar to that found  in absence of super-coiling. For $0< \kappa<1$   the anchoring 
plate confines the molecular chain in  the $ z>0 $ half space.

\appendix*
\section {Calibration of the stretching force from  elongation and  transverse
fluctuations  dsDNA   measurements} 

We would like to use partition function transformation properties under space
rotation to derive the Strick {\it et al.} \cite{strick} formula, within the WLC model in the limit
where the molecule contour  $ L$ is larger than the
persistence length $A$. This appendix  is an updating of an unpublished internal note of the author
\cite{stformCB}.
 
The thermal averaging over the various configurations of the molecular chain is
performed   here  within  the continuous version of the WLC model,
 using  a technique inspired by quantum mechanics.
 It leads to the following expression  for the partition function:
\begin{equation}
Z(\hat \vt_0,\hat \vt_1, \vF, L) = \Sigma_n  \Psi_n(\hat \vt_0\cdot \hat
\vz)\Psi_n(\hat
\vt_1\cdot \hat \vz) \exp -{\left( \epsilon_n(\alpha)\frac{L}{A}\right)}\,,
\label{qmwlc}
\end{equation}
where $\hat \vt_0, \hat \vt_1$ are the unit tangent vectors along the chain at the two
ends $s_0=0$ and $s_1=L$,
 $\epsilon(\alpha) $ and
$\Psi_n(\hat \vt\cdot \vz)$ are respectively  eigen-values and eigen-functions of the WLC
Hamiltonian $H_{WLC}$ \cite{QMvsSM}. The partition function written above is clearly invariant
upon simultaneous rotations of $\hat \vt_0, \hat \vt_1$ and $\vF$ but it is modified
by rotating $\vF$ while keeping $\hat \vt_0, \hat \vt_1$ fixed or {\it vice versa}. If
the experiments are performed under the condition $L\gg A$.  it is legitimate to make
two simplifications in the right hand side of Eq. (\ref{qmwlc}):

i) The sum over $n$ is dominated by the ground state contribution $(n=0)$, since the
excited state contribution is  strongly suppressed by a factor of the order of
$\exp{-( \Delta \epsilon_1 L/A)}$, where $\Delta \epsilon_1 =
\epsilon_1-\epsilon_0 \sim 1$.

ii) Keeping only the ground state term, the logarithm of the partition function reads
as follows:
\begin{equation}
\ln{Z(\hat \vt_0, \hat \vt_1, \vF,L) = -\frac{L}{A }\epsilon_0(\alpha) \left(1 +
O(\frac{L}{A})\right)}.
\label{thermolim}
\end{equation}
The term $O(\frac{L}{A})$ comes from the prefactor involving logarithms of the
ground state wave function.

{\it So we can conclude that the free energy of the molecule stays invariant
when one rotates the force $\vF$, while maintaining fixed the two end tangent
vectors $\hat \vt_0, \hat \vt_1$, provided finite size effects of the order of $\frac
{A}{L}$ are ignored }.

We are now going to exploit this result by using   the path integral form of the
partition function:
\begin{equation}
Z(\hat \vt_0, \hat \vt_1,\vF,L) = \int {\cal D}(\hat \vt) \exp{-\left( \frac{E_{bend}
+E_{stretch}}{k_B T} \right)}\,,
\end{equation} 
where ${\cal D}(\hat \vt)$ is the path integral measure and $E_{bend}$ is the elastic
energy describing the resistance against bending. The stretching energy
$E_{stretch}$ is the potential energy associated with the uniform force $\vF$ 
applied to the free end of the DNA chain:
\begin{equation}
E_{stretch}=-\vr (L)\cdot \vF = -\int_0^L ds \; \hat \vt (s) \cdot \vF \,, 
\end{equation}
where $\vr (L) $ is the coordinate of the end point. Instead of taking as usual $\vF$
 along the $z$-axis, let us  apply to $\vF$  a rotation   ${\cal R }_y(\delta) $  of angle $\delta $ about the
$y$-axis, leaving unchanged $\hat t_0 $ and $\hat t_1$. The stretching energy
reads, then , as follows:
\begin{equation}
E_{stretch}(\delta)=-(  z (L) \cos{\delta} + x (L) \sin{\delta}) F.
\end{equation} 
By writing that the Taylor expansion of $\ln{Z(\hat \vt_0, \hat \vt_1, \vF, L)}$ in the
vicinity  of $ \delta = 0$ vanishes term by term,  we shall obtain a set of linear
relations involving the moments of $x(L)$ and $y(L)$. Each moment is weighted by
powers of $f = \frac{F}{k_B T}$ such that the relations involve dimensionless
quantities. In fact, the relevant relation is obtained by writing:
\bea
\lim_{\delta \to 0} \frac{1}{Z(\delta)} \frac{\partial^2 Z(\delta)}{\partial\delta^2} 
&= & \lim_{\delta \to 0} \bra \,\exp \(( \frac{  E_{stretch}(\delta)  }{k_B T} \))
 \frac{\partial ^2}{\partial \delta^2} \exp-\((  \frac{E_{stretch}(\delta)}{k_B T}\)) \ket  \nonumber \\
 & = &   f^2 \bra x^2 ( L) \ket  - f \bra   z(L) \ket , 
\label{stalform}
\eea
where the r.h.s. has to be understood as a thermal average.
If one uses   the rotation  invariant partition function   given by eq. (\ref{thermolim}),  
  the l.h.s of the  above  equation vanishes, up to  corrections of the order of $\frac{A}{L}$. One   arrives
then  immediately  at  the formula of Strick {\it and al.} \cite{strick}
which  gives the stretching force F  in term of the DNA
elongation and the transverse fluctuations of the free end:
\begin{equation}
F = \frac{\bra z(L) \ket \,k_B T }{ \bra x^2 (L) \ket } \left( 1 + O(\frac{A}{L}) \right). 
\label{STform}
\end{equation}
Note that the above derivation does not use any small angle approximation.

We  shall try to   estimate  the  correcting terms under conditions such that
$\frac{A}{L}$ is no longer negligible, say $\frac{A}{L} \sim 0.2$, while
$\exp{ (-\frac{L}{A})} \sim 0.007$ is still very small. The simplification i)
is still legitimate,  so we can use for the partition function the following approximate
form:
\begin{equation}
Z(\hat \vt_0,\hat \vt_1, \vF, L) \simeq \Psi_0(\hat \vt_0\cdot \hat \vz)  \Psi_0(\hat
\vt_1\cdot \hat \vz) \exp{\left(-\epsilon_0(\alpha) \frac{L}{A}\right)}. 
\label{partf}
\end{equation}
If  we apply the  rotation   ${\cal R }_y(\delta) $ to  the force, 
the partition  function of  eq. (\ref{partf}) is then 
$ \propto  \Psi_0\(( \cos( \theta_0 -\delta) \))\  \Psi_0\(( \cos( \theta_1 -\delta)  \)). $
The computation  of   the finite size correcting term:
 $ \Delta_{f.s}( \theta_0,\theta_1) =      \lim_{\delta \to 0}\frac{1}{Z(\delta)} \frac{\partial^2Z(\delta)}{\partial\delta^2}$ 
is straightforward, but the final result is rather lengthy.
  The calibration formula corrected for finite size effects, valid    for any  value of  $ \theta_0$  and $\theta_1  $,
is then readily obtained  from  eq.(\ref {stalform}):
\be
F= \frac{\left<z(L)\right>k_B T}{\left< x^2(L)\right> } \(( 1 +  \Delta_{f.s}( \theta_0,\theta_1) \, \frac{A}{\bra z(L) \ket \,\alpha}
+O(\frac{A^2}{L^2}) \)). 
\label{fnsizcor}
\ee 
 With the above writing,   the finite size   correction  is proportional    to 
$ A/( \bra z(L) \ket \,  \alpha)$. We  are going now  to estimate the  prefactor $   \Delta_{f.s}( \theta_0,\theta_1) $.  
  It is of interest to quote  the result  in
the simple case $\theta_1= \theta_0=0$ :
\begin{equation}
 \Delta_{f.s}( 0,0) =   \lim_{\delta \to 0}\frac{1}{Z(\delta)} \frac{\partial^2
Z(\delta)}{\partial\delta^2}= -2 \frac{\Psi^{\prime}_0(1)}{\Psi_0(1)}.
\label{simplecor}
\end{equation}

 Let us first consider   ``large"  force values :  $\alpha  \gg 1 $,  remembering that $ \alpha =4$  corresponds to $F =0.3$ pN !
 An  approximate ground state wave
function can be easily derived, together with the corresponding  eigen-energy:
 $$ \Psi_0(\cos{\theta}) \propto \exp{- \frac{1}{2} (\sqrt{\alpha} \, \theta^2)}\; \;\text{and} \; \; \epsilon_0  (\alpha)= -\alpha+
\sqrt{\alpha}\,.$$   

It leads immediately to the relative  extension: $\left< z(L)\right> /L
\simeq 1- \frac{1}{2 \sqrt{\alpha}}$ ( this value is very close to the  
 exact one when $ \alpha  \geq 4$). The correcting term  $\Delta_{f.s}(\theta_0,\theta_1) $  
 can be computed  easily   for arbitrary  $ \theta_i $  angles:
\be 
   \Delta_{f.s}( \theta_0,\theta_1) =
-  2 \sqrt{\alpha }+ \alpha  \left(\theta _0+\theta _1\right)^2.
\ee
 The above formula shows clearly  that something has to be said about the angles $ \theta _0$  and   $ \theta _1$. The angle  $ \theta _0$  
gives the direction of the initial strand, having a length  of about $0.1 \, A$,  sticking out from the anchoring plate. The angle is 
partly  determined, first,  by the biological gluing process,  second, by the fluctuating  tension  force induced by 
 the rest of the chain. If the latter mechanism were  the dominant one, a thermal average would have to be  performed, using the  probability 
distribution $ \propto \Psi_0 ( \cos \theta _0 )$. A similar analysis holds for the terminal strand, sticking into the magnetic bead.
Computing the thermal average  $ \bra (\left(\theta _0+\theta _1\right)^2 \ket = 2 \bra \left(\theta _0\right)^2 \ket = 2 \sqrt{\alpha}$, 
one gets $\Delta_{f.s}( \theta_0,\theta_1)=0 $. ( This result turns out    to be  valid for arbitrary forces. It hinges  
 upon the fact that   thermal  averaging over  $ \theta _0$  and   $ \theta _1$ and the  second derivative 
$ \frac{\partial ^2}{\partial \delta^2}$  are commuting operators.The latter  statement  is not totally trivial since the 
wave function $  \Psi_0 ( \cos \theta  )$ is restricted to the finite interval  $   0 \leq \theta  \leq\pi $.) 
  The above considerations    lead  to the following approximate bounds for the finite size correction:
\begin{equation}
\alpha \gg 1 \Rightarrow \;  -2 \frac{A}{L} \frac{1}{\sqrt{\alpha}} \lesssim F\, \frac{\left<x^2(L)\right>}{\left< z(L)\right> k_B T}
-1 \lesssim  0 .
\end{equation}
Let us apply  the above result to a situation considered in section II,  involving a  ``short" molecular chain  
$ A/L = 1/12$  together with a stretching force  $ \alpha = 4  \rightarrow  F= 0. 3$ pN . The large   $ \alpha $ asymptotic 
formulae give already a fair approximation. One sees   immediately  that   the finite size correction 
  stays  below the $ 10\,  \% $ level, when $ \alpha \geq 4$.

In the small $\alpha $ limit,  the ground state wave function is obtained by a first
order perturbation calculation:
$$\Psi_0(\cos{\theta})= \frac{1}{\sqrt{2}}(1+ \alpha \cos{\theta}+ O(\alpha^2)).$$
Performing some simple algebra, one gets:
 $$ \frac {\left<z(L)\right>}{L}= \frac{2}{3}\, \alpha \; \; \text{ and} \; \;\Delta_{f.s}( \theta_0,\theta_1) =
- \alpha \((\cos (\theta_0)+\cos(\theta_1) \)) +  O(\alpha^2). $$
 In  the present low force  limit  the initial and final angle, $\theta_0 \;  \text{and} \; \theta_1$,  are weakly affected by 
the tension of the chain. To take into account the spatial constraints associated with the anchoring plate and the magnetic bead 
we  impose the restriction $ 0 \leq \theta_i \leq \pi/2$. We arrive in this way at the following bounds:
    $$\alpha\ll 1 \Rightarrow -\frac{3}{\alpha} \frac{A}{L}  \leq  F\, \frac{\left<x^2(L)\right>}{\left< z(L)\right> k_B T} -1
    \leq 0 .$$ 
 Let us take as a typical low force  case $ \alpha= 1/3  \Rightarrow  F \simeq  1/40$ pN.
  Using the relation   $ \bra\bra \cos \theta \ket \ket = \bra  z(L)  \ket /L= 2/9 $,  one gets  for the average angle 
between the running tangent vector and the force direction a value of about 1.3  rad.
If  one takes $ A/L \simeq 1/10 $, 
  the correction can  indeed  reach values  of the  order of unity. However,  if one deals instead with  a ``long"  molecule with  $ L =300\, A$, 
as in the experiments analyzed in  ref. \cite{bouchiat99}, one gets $ 3\, A /( \alpha \, L )= 3 \times 10^{-2}$.
The formula (\ref{STform}) is  then rather accurate, despite the fact that
 it was originally derived within a small angle approximation.


\begin{thebibliography}{30}
\bibitem{WLCelong} In  ref. \cite{bouchiat99} the  relative elongation $ u= \bra z(N) \ket /L $  versus 
$ \alpha = F \,A /(k_B \,T )$  is given by: $$ \alpha =  {\cal F }(u) = u+\frac{1}{4 (1-u)^2}-\frac{1}{4} +{\cal P }(u), $$
where the polynomial $ {\cal P }(u) $  provides an interpolation of the   numerical data  obtained by solving accurately
the Schr\" odinger equation associated with the WLC model in absence of spatial constraints:
$$ {\cal P }(u) =   -14.17718\, u^7+39.49944\,u^6-38.87607 u^5+  
         16.07497 u^4-2.737418\, u^3-0.5164228 u^2. $$
\bibitem{QMvsSM}  The  Hamiltonian $ H_{WLC}$  is invariant
 under  rotations around  the force direction $ \hat\vz $. As a consequence the   eigen-functions of   $ H_{WLC}$
have the general form: $ \Psi_{n,m}(\hat\vt \cdot \hat \vz) \, \exp( - i \, m \, \phi )$  where $ \phi $ is the azimuthal angle of
 $ \hat\vt$ around $ \hat \vz$. Some readers may wonder why the wave functions with $ m \ne 0$  do not appear  in the
partition function. The reason is that, in contrast with   the case of  Quantum Mechanics, the wave function itself  has   to be interpreted as  a
probability  distribution.  The choice $m=0 $ is  then dictated  by this  physical requirement.
\end{thebibliography}
\end{document}